# Gender Issues in Computer Science: Lessons Learnt and Reflections for the Future

Letizia Jaccheri
Norwegian University of Science and Technology
Department of Computer Science
Trondheim, Norway
letizia.jaccheri@ntnu.no

Cristina Pereira
Informatics Europe
Zurich, Switzerland
cristina.pereira@informatics-europe.org

Swetlana Fast
Norwegian University of Science and Technology
Faculty of Information Technology and Electrical Engineering
Trondheim, Norway
swetlana.fast@ntnu.no

*Abstract*—Women are underrepresented in Computer Science disciplines at all levels, from undergraduate and graduate studies to participation and leadership in academia and industry. Increasing female representation in the field is a grand challenge for academics, policymakers, and society. Although the problem has been addressed for many years, progress has been difficult to be measured and compared across countries and institutions, and has been invariably slow, despite all the momentum and impulse for change taking place across several countries. Therefore, it is important to reflect on knowledge, experiences, successes, and challenges of existing policies, initiatives and interventions. The main goal of this paper is to provide an overview of several initiatives, studies, projects, and their outcomes. It contributes to building a body of knowledge about gender aspects in several areas: research, education, projects, networks and resources. This paper is mainly based on discussions in working groups and the material collected for and during a series of talks on the topic held by the first author and by feedback received by the community. This paper provides the academic community, policymakers, industry and other stakeholders with numerous examples of best practices, as well as studies and recommendations on how to address key challenges about attracting, retaining, encouraging, and inspiring women to pursue a career in Computer Science. Future work should address the issue in a systematic and research based way.

## I. INTRODUCTION

In recent years, women's underrepresentation in science, technology, engineering, and mathematics (STEM) fields has captured the attention in public, academic, and policy circles. While the gender gap in STEM is widely discussed and recognized, its relative size among different sciences, technology and engineering are less evident. Computer Science (Informatics, Computer Engineering, Computing, Information Technology (IT)) is, unfortunately, one of the most heavily affected areas. Areas such as Chemistry and Biology have a much more balanced gender distribution, whereas it is predominantly in Computer Science (CS), Engineering, and Physics that women are under represented. This strong imbalance is long enduring, and no significant progress has been observed in the past years, either in Europe [1] or the US [2], [3] An analysis of CS literature [3] estimates that the gender gap in CS research (parity between the number of male and female authors) will not close for at least 100 years if specific measures are not taken.

A temporal analysis of the data shows that, on average, no significant progress in female participation in CS higher education has been observed over the past ten years in Europe [4]. The same is true for the US, as shown in [5], gathering data on college students for four decades, which highlights a persistent, sizable underrepresentation of women in CS in the US.

Beyond academia, the profession also inherits the male-dominated student population. Women are strongly underrepresented among ICT specialists in all EU Member States, a striking contrast with total employment, where the genders are broadly balanced.

Despite the clear negative impact and consequences of a strong gender unbalanced environment, unfortunately, the fight for gender balance and equality in CS is seen and dealt with as a women's problem. Projects, programs, actions, strategies are invariably led by highly motivated and achieving women who volunteer their time to establish a more equal environment and pave the way for the new generations of female scientists. Going beyond their daily work, they are responsible for the effort, and comparatively more moderated funding, that has been spent on the efforts for gender equality in CS.

Increasing representation of women in CS at all levels, from undergraduate and graduate studies to participation and leadership in academic and industry positions, is a grand challenge for academics, policymakers and for our society. While research has shed light on the issue, there is limited evidence on effective solutions and on what works. This paper is based on [6] and summarizes knowledge collected by several workshops [7]. This paper addresses gender issues in CS from five perspectives: gender issues in the society, research, education, projects, networks and resources (see Fig. 1). The paper casts light on the current status-quo in each of these areas and provides an overview over the best practices that help to tackle gender imbalance in the field. It provides the academic community, policymakers, industry and other stakeholders with outstanding examples on how to address the following key challenges:

(i) How to have more girls choosing CS as their higher education studies and profession;

(ii) How to retain female students and assure they finish their studies and start successful careers in the field;

(iii) How to encourage more female Ph.D. and postdoctoral researchers to remain in the academic career and apply for professorships in CS departments;

(iv) How to support and inspire young women in their careers and help them to overcome the main hurdles that prevent women from reaching senior positions in industry and public sector.

(v) Which communication and dissemination strategy to adopt in this field.

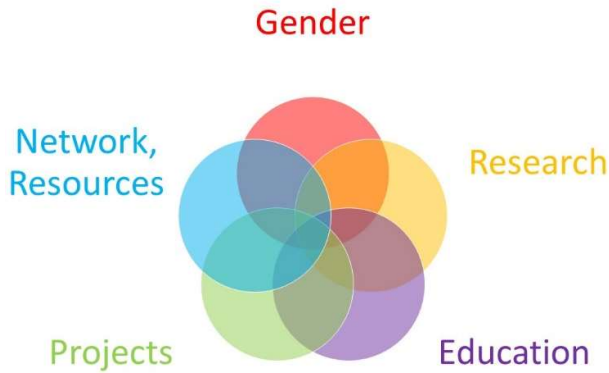

Fig. 1. Five lenses: Gender, Research, Education, Network & Resources, and Projects

## II. Gender Issues in the society

Gender Equality is one of the United Nations' goals (Goal 5) towards a peaceful, prosperous and sustainable development(Fig. 2). Stopping and preventing discrimination against women is essential to assure gender equality as a fundamental human right. Besides this, empowering women fosters economic growth and development. Over the last decades, there has been worldwide progress on many societal developments such as schooling and marriage rights or access to the leading positions in governmental bodies, education and industry.

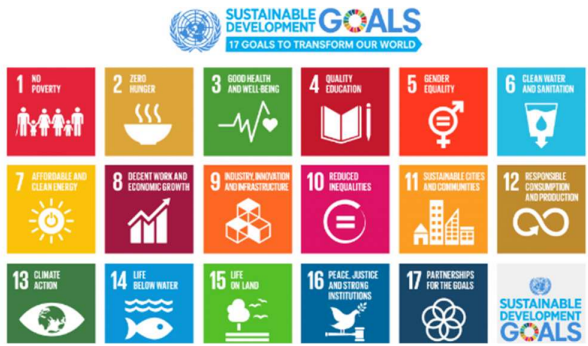

Fig. 2. Goal no. 5 Gender Equality is one of the United Nations' 17 sustainable development goals

Across Europe, gender equality in all aspects of social, political, and cultural life, including education research, and innovation, is approached as a matter of social justice and fairness. Toward the end of achieving gender equality in STEM at large, targeted measures have been developed and actions undertaken at the national and European levels. The European Union's Strategy for Gender Equality in STEM is targeting the labour market and research with three main objectives: gender equality in careers, gender balance in decision-making bodies, and integration of the gender dimension in Research and Innovation (R&I) [8]. The European Commission has also outlined a strategy to encourage and empower women to play a more active role in the digital age.

The Women in Digital (WID) initiative [9] also includes the WID Scoreboards created to monitor women's participation in the digital economy in all Member States. On Digital Day 2019, 26 EU Member States and Norway have signed a declaration to encourage women to play an active and prominent role in the digital and technology sectors, committing to work closely with the public and private sectors and civil society to achieve equality in tech. The declaration was also signed by numerous CEOs and leaders of major EU and international companies.

## III. Research

### A. Technology With Gender Biases

It has been observed that the tech industry's dominantly male workforce leads to the creation and development of technology and systems including gender bias. Many examples are found on the EU Report of the Expert Group "Innovation through Gender" and website of the international project on Gendered Innovations [10]:

- smartphone voice assistants (Android's Cortana, Apple's Siri, Microsoft's Alexa) that are all female with noticeable submissive personalities and ill-equipped to respond to user requests regarding crises that predominantly affect women (e.g., sexual assault) [11];
- activity trackers that fail to measure steps in the, predominantly female, activity of pushing a stroller;
- transport networks that ignore the so-called "mobility of care"
- AI recruiting technology developed trained predominantly on men's résumés

According to the first author who is a software engineering researcher, user diversity is highly important since the requirements, interaction and communication channels, and in the end also products and services can be influenced by the biases of the developers. In software engineering and AI, where computers are "trained" how to learn, it becomes even more important that a broad variety of samples must be involved, including different genders, ethnicity, age and health conditions.

### B. Gender And Story Telling

Gender stereotyping in child development and education is a known issue, but until now little attention has been given to the design for change, or at least attenuate, stereotypical thinking. In the research documented in [12]12 to which the first author has participated, a Digital Story Telling (DST) tool has been developed and studied to support children in their awareness of negative gender stereotypes. Following a participatory design-inspired approach, the study involved 43 participants, children and adults (teachers and experts in the domain), in three workshops with the purpose of exploring this design space. The study proposes concepts to guide the design of a DST tool as an outcome. The workflow and toolbox used during the process were instances of an approach that could be replicated in other contexts and/or to challenge other types of stereotypes. The main contribution of this research is towards the design for change with and for children.

*C. Software and games for equality*

The objective of exposing girls to CS as a career option has led to research directed towards gaming activities for girls. These activities include both game play and game design. The status of research has been provided through a systematic literature review to which the first author has participated. The study investigates the relation between the various game playing or designing activities and their impact on girls' perception of CS as a career choice. The study presents the design consideration for the games and related activities to potentially improve the perception of girls towards a CS career [13]. The review identified 25 papers to distil a common understanding of the state-of-the-art. Specifically, the paper investigated the effects that the game play/design activities had on girls' perception about CS; and what are the key design factors to be kept in mind while designing a serious game to improve girls' perception about CS. The results of this systematic literature review showed that game playing or designing could improve how girls perceive having a career in CS. The key aspects that such activities require are personalizing, opportunity for collaboration and the presence of a female lead character [14].

The first author has leaded a project [15] that has the goal of providing knowledge on how to develop software that help to address the goal of preventing child marriage. Fig. 3 illustrates the interview sessions with girls in Malawi answering questions about child marriage.

Girls

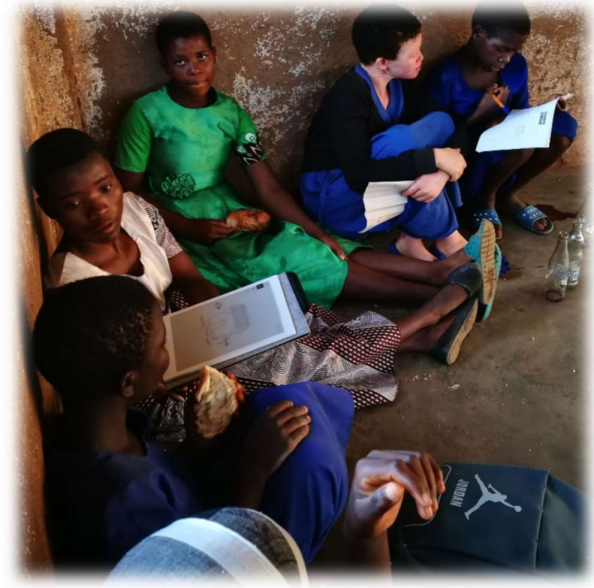

Fig. 3. Interview session with girls in Malawi.

## IV. HIGHER EDUCATION

*A. The CS Gender Gap in Europe*

Higher education statistics for numerous European countries, collected along the past 10 years, corroborates that the strong female underrepresentation in higher education in CS in Europe is a long-standing problem.

The investigation on the female participation revealed that at the Bachelor level, in Austria, Belgium, Denmark, Finland, Germany, Ireland, Italy, Latvia, Lithuania, the Netherlands, Poland, Spain, Switzerland, and the UK, 80% or more of the students enrolling or graduating in Informatics Bachelor programs are male. Similarly, In Bulgaria, Greece, Romania, and Estonia a slightly narrower gap exists, however, women do not represent more than 30% of the bachelor students [1].

The share of women in higher studies is still very low for MS and Ph.D. with female students' enrolment of around 20% and 25% respectively. At the Master level female participation increases in some countries, over 35% of the Master graduates in Bulgaria, Romania and Greece, and around 30% in the UK, Estonia, Ireland, and Latvia, but decreases in others, not surpassing 20% of the Master graduates in Austria, Belgium, Czech Republic, Germany, Italy, Lithuania, the Netherlands, Poland, Portugal, Spain, Switzerland.

At the Ph.D. level, except for Bulgaria, Romania, Estonia, Turkey, all other countries have less than 25% of women graduating from CS Ph.D. programs, corresponding in some cases to less than a handful of women, as the total number of Ph.D. graduates in many countries is quite small.

Research on CS education has focused on gender differences and there is high interest in increasing female participation in CS. A previous study reviewed important factors that influence CS university students' decision to complete their studies in CS (retention), and empirically examines how these factors differ for males and females. To this end, it was identified cognitive and non-cognitive gains, cognitive and affective engagement, motivation to study, and three different barriers as critical factors in CS students' choices. The authors tested these factors on 236 Norwegian CS students in order to find any potential gender differences. The findings indicated a gender difference for cognitive gains, affective engagement, motivation to study, and satisfaction with learning effectiveness. On the other hand, no difference was found for non-cognitive gains, cognitive engagement, personal values/teaching quality, and students' intention to dropout from their studies. The study concluded with implications that will guide both research and practice towards a better understanding of both male and female CS students [16].

*B. Scarcity of Role Models*

Only a small proportion of the few women graduating with a Ph.D. in CS will pursue an academic career and fewer will progress to the highest academic ranks of associate and full professor. Similarly, to other STEM areas, in CS the pipeline is leaking, and the ceiling made of strong glass. Data for CS only is not available at the European level, however, a loose parallel can be drawn from data for Engineering and Technology where on average, in the whole of Europe, women take less than 15% of the full professor positions [1]. The very low number of women reaching senior academic positions results in a scarcity of successful female role models to influence the new generations. To be a distinct minority in academia also results in the overload of invitations and requests (conferences talks, committees, administrative department roles, etc.), which penalizes women's careers, impacting negatively their research productivity, their work-life balance, their personal life, and health.

*C. The skills And Talent Gap In The ICT Professions*

Similarly, to the situation in higher education, figures show that a majority of ICT specialists employed in the EU are men. An EU study from 2016 showed and average across Europe of 83.3% of male ICT specialists. This was the case in every EU Member State, the highest shares of male ICT specialists were observed in Slovakia (90.8%), Czech Republic (88.8%), Malta (88.3%), Greece (87.3%), Hungary (86.9%) and Croatia (86.7%), while Bulgaria (69.8%), Romania (73.7%), Latvia and Lithuania (both 75.2%) recorded the lowest [17].

The lack of women is among the core reasons for the huge skills and talent gap existing between the number of graduates in higher education institutions and the number of job positions available in the ICT Industry in Europe.

Currently, an average of 53% of European employers says they face difficulties in finding the right people with the right qualifications. The highest percentages were recorded in the Czech Republic (79%), Austria (78%), Malta (73%), Luxembourg (71%), the Netherlands (69%), Slovenia (65%), Germany (64%), Denmark (61%), Belgium, Hungary, Slovakia, Finland, and Sweden (60%) [18]. Hundreds of thousands of vacancies for ICT professionals in Europe remain unfilled and this situation tends to aggravate with our society moving to a pervasively digitalised world and the unprecedented developments in technology. The talent gap in ICT is one of the most serious threats to the economic development of Europe.

V. PROJECTS

In Europe, several universities have set up projects, internal policies, and strategies specifically to increase the number and retention of female students and scientific employees in CS. Table 1 provides examples of such initiatives. These have not to be considered as an exhaustive list nor as result of a systematic search.

TABLE 1. GENDER BALANCE PROJECTS AT UNIVERSITIES

| Title | Ref. | Where | Description |
|---|---|---|---|
| The Bamberg CS30 Strategy | [19] | Faculty of Information Systems and Applied Computer Sciences, University of Bamberg Germany | Started in 2005 and aims at reaching a female/male ratio of at least 30% across all CS programs. The number of women enrolling in first-year CS studies has been increasing since 2013 and reached 37% in 2017, establishing a new record in Germany. |
| The Girl Project Ada | [14] | Faculty of Information Technology and Electrical Engineering, Norwegian University of Science and Technology (NTNU)Norway | Started in 1997 and aims at recruiting more girls to the STEM studies and prevent dropouts. The female share of entrants in STEM studies has, on average across different programs, almost doubled, going over 25% in the CS program in 2017, and from 4% in 2004 to 36% in 2019 in STEM disciplines. Ada implements various measures, e.g. invite girls from high school from all over the country, personal meetings with role models, technology days, mountain hiking tours, coding events, PhD Parties (see Fig. 4). |
| CS4All | [20] | School of Computing, TU Dublin Ireland | Started in 2012 and aims at increasing the number of female students coming to CS undergraduate programmes and reducing the numbers of those failing to progress in the critical first year. The females' share of enrolled students in a new CS Bachelor Program, with a strong emphasis on Internationalisation and Globalisation (22%) is double the one of the standard CS in the same period. Retention has been strongly improved, particularly for first-year students with an average 89% progression from year 1 to year 2 (the most critical), now the highest progression rates for CS in Ireland. |
| Irène Curie Fellowship | [21] | TU Eindhoven, The Netherlands | This program aims at attracting talented female scientists. Irène Curie Fellows receive a tenure track position as assistant professor of 5 years and after further 4 years the possibility to apply as associate of full professor. it offers a tailor-made career proposal, opportunity to establish an independent research program; substantial start-up package to kick-off career; training programs for academic leadership and the university teaching qualification; mentoring scheme, valuable networks, flexible work schedules, family friendly initiatives. |
| Genie, Gender Initiative for Excellence | [22] | Chalmer, Sweden | Through concrete changes of the academic culture, system and procedures together with strategic recruitments, Genie aims towards equal gender balance within the faculty. The university invests 300 Mil. SEK over ten years on gender equality work. The initiative is funded by the Chalmers University of Technology Foundation. It was launched on 1 January 2019 and runs until 2028; the largest individual investment in gender equality made by any university. |
| IDUN From PhD to Professor, Gender Balance in Top Research Positions | [23] | NTNU, Norway | IDUN aims at increasing the number of female scientists in top scientific positions. Budget: 1 Mio. Euro, duration: August 2019-July 2022. The project has recruited nine female adjunct professors acting as mentors and role models for 35 early-career researchers at faculty's seven departments. IDUN Scientific Mentoring Program focuses on networking, group work on a specific research topic, proposal writing and career planning activities. |

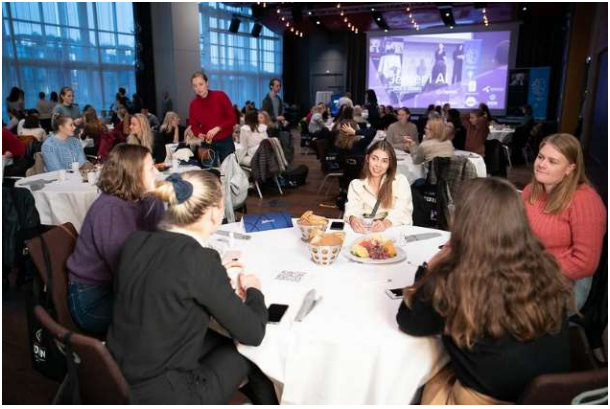

Fig. 4. Students meet their female role models from industry and academia at Girls in AI 2020 event [14]

Many more Universities have individual projects or fellowships (involving directly or indirectly CS Departments), aiming at improving gender balance. A few examples, involving substantial funding, include: The Irène Curie Fellowship at TU Eindhoven, the Gender Initiative for Excellence at the Chalmers University of Technology, and the IDUN project at NTNU. These are listed in Table 2.

Frequently applied measures are mentoring programs for women scientists, equal representation in the recruitment processes and work environment support. Fig. 5 shows an example from the IDUN project (for which the first author is project leader) with the main building blocks.

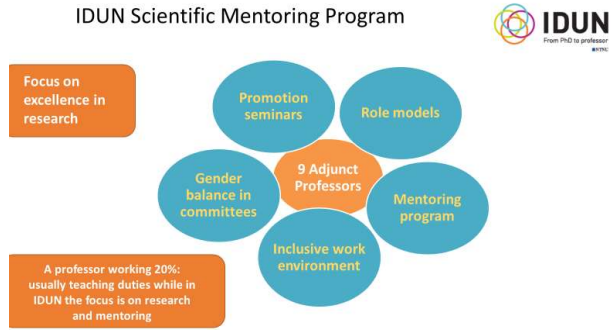

Fig. 5. An example from Norway, IDUN Scientific Mentoring Program at NTNU with nine scientific mentors and focus on excellence in research

Outside Europe, some remarkably successful examples at the university level can be found in the US. The most famous examples are Harvey Mudd College [24] and Carnegie Mellon [25] where in the past decade gender parity has been achieved in CS entrants and graduates. Although inspiring, these efforts remain isolated and proved difficult to escalate to more institutions and to improve the national statistics.

Similarly, to most of the STEM fields, measures aiming at gender balance in CS is invariably driven by women. The focus on integration should be on male colleagues and stakeholders, to raise their awareness and understanding of the threat imposed by the low numbers of women in the field.

## VI. NETWORK, RESOURCES

Compared to the US, Europe still lacks behind on what regards the amount of funding, successful examples and the level of organization of the community. Organizations and groups such as AnitaB.org, ACM-W, CRA Women, National Center for Women & Information Technology, IEEE Women in Computing Committee, Association for Women in Computing, Girls Who Code, in partnership with industry have established a thriving community empowered to inspire and encourage the new generations and to support the careers women in CS. The most spectacular example of this community is the Grace Hopper Celebration, which in 2018 gathered over 20,000 attendees from over 78 countries, almost all women, at all stages in CS studies and career, providing an invaluable opportunity for women to find inspiration, networking, and strategies to thrive in their careers. Some significant events are shown in Table 2.

TABLE 2. CONFERENCES

| Title | Ref. | Where | Description |
|---|---|---|---|
| Grace Hopper Celebration | [26] | US | Created in in 1994 and inspired by the legacy of Admiral Grace Murray Hopper, the AnitaB.org flagship event Grace Hopper Celebration brings the research and career interests of women in computing to the forefront. |
| ACM womENcourage | [27] | International, 2019 in Rome, Italy; 2020 in Baku, Azerbaijan (online) | ACM Celebration of Women in Computing; supporting, celebrating and advocating for Women in Computing |
| IEEE Women in Engineering International Leadership Conference (IEEE WIE ILC) | [28] | International | Launched in 2012, provides professional women in technology (in industry, academia, or government) the opportunity to create communities for innovation, knowledge sharing and mutual support. It focuses on providing professional development for mid-level and senior level women. |

Table 3 provides an overview of initiatives to improve diversity and integration in CS:

Networking among actors involved in current European efforts from the outset, not only helps intensify the pace and effectiveness by drawing on existing information, but also stimulate experimentation and conceptual adaptation by understanding differing approaches. Networking also ensures that the work done by these projects is afforded further sustainability and an opportunity to be built upon, by integrating their efforts, ensuring these strides and collaborations are consolidated to create a more holistic, practicable and evaluated approach that can inform local, regional and EU practice and policy regarding the gender balance in CS.

In Table 4 we list additional examples of relevant activities, white papers and reports.

TABLE 3. NETWORKS AND PROJECTS

| Title | Ref. | Where | Description |
|---|---|---|---|
| Informatics Europe WIRE | [7] | Europe | Promote actions that help improve gender balance at all stages of the career path in Informatics |
| The ACM Council on Women in Computing in Europe (ACM-WE) | [29] | Europe | Transform European professional and scholarly landscape where women are supported and inspired to pursue their dreams and ambitions to find fulfilment in the computing field. |
| European Centre for Women and Technology (ECWT) | [30] | Europe | European multi-stakeholder partnership of more than 130 organizations representing high-level expertise in women and technology development from government, business, academia and non-profit sectors working together to measurably and significantly increase the number of girls and women in technology and ICT. |
| CEPIS Women in ICT Task Force | [31] | Europe | It pools the expertise from its Member Societies to share best practices and develop pan-European activities to increase women's participation in the ICT profession. |
| The EQUALS Global Partnership for Gender Equality in the Digital Age | [32] | International | An international multi-stakeholder initiative dedicated to promoting gender balance in the technology sector. |
| Women4IT | [33] | EEA and Norway | Aims at developing the digital competences of young women who are at risk of exclusion from the labour market, by improving their employability |
| COST Action CA19122 EUGAIN European Network For Gender Balance in Informatics | [34] | Europe | EUGAIN aims to improve gender balance in CS through the creation and strengthening of a multi-cultural European network of 37 countries and academics on the forefront of the gender balance efforts in their countries, institutions and research communities. The project started in October 2020 and has a duration of 4 years. |
| LIBRA | [35] | Europe | Unifying innovative efforts of European research centres to achieve gender equality in academia; aims at increasing the representation and participation of women in leadership positions in life sciences |
| GEECCO, Gender Equality in Engineering through Communication and Commitment | [36] | Europe | Aims at developing dos and don'ts while degenderizing STEM |
| PLOTINA, Promoting gender balance and inclusion in research, innovation and training | [37] | Europe | Works on contrasting unconscious gender biases in the evaluation and recruitment of professors and researchers |
| GENDERACTION, Gender equality in the ERA Community To Innovate policy implementation. Coordination and Support Action | [38] | Europe | Supports the implementation of gender equality in the European Research Area |
| Gesellschaft für Informatik in Germany, Fachgruppe für Frauen in Informatik | [39] | Germany | Advocate equal participation of women in IT, equal opportunities in professional life, in terms of salary, career and working conditions, and the greater awareness of the high level of competence of women computer scientists in the professional world and in the public. |
| EQUAL-IST Project | [40] | Europe and CIS (Commonwealth of Independent States) | Aims at introducing structural changes in Information Sciences and Technology research institutions to enhance gender equality. |

TABLE 4. OTHER RESOURCES

| Title | Ref. | Where | Description |
|---|---|---|---|
| I'd blush if I could | [11] | International | Unesco report about gender gap in ICT education |
| Innovation through Gender | [10] | EU, International | Report for scientists and engineers, provides practical methods for gender analysis, and to develop case studies as concrete illustrations of how gender analysis lead to new ideas and excellence in research. |
| SHE Figures | [1] | Europe | Report that gives evidence to the progress achieved over the past years in gender balance through a wide range of actions and policies. |
| The researcher journey through a gender lens | [41] | Europe | Report that examines research participation, career progression and perceptions across the European Union and 15 countries in 26 subject areas. |
| Informatics Higher Education Data Portal | [4] | Europe | Online repository that provides a picture of the state of Informatics higher education in Europe with statistics for Bachelor, Master and PhD level students (first year, enrolment and graduates). It include gender distribution. |
| Gender Equality in Academia and Research Tool (GEAR) | [42] | Europe | A tool provided by the European Institute for gender equality plans for academic and research organizations, including recruitment, selection, and career progression support. |
| Women in Digital. Digital Single Market Policy | [9] | Europe | Policy report about the European Commission's initiatives encourage and empower women to play a more active role in the digital age. |

## VII. FUTURE WORK

This paper is a collage of information collected during workshops, presentations, and discussion and should not be regarded as a complete nor validated framework. To consolidate this work, it is necessary to progress into several directions. Specifically, the community should:

1. Set up new research projects and follow up existing ones that seek understanding of practices, interventions and different kinds of policies and analyse if they can be applied and potentially work across different institutions, countries, cultures.

2. Plan dedicated activities that aim at transferring the existing knowledge to different stakeholders. To

improve knowledge development, dissemination, and visibility of the existing actions it is crucial to involve the following stakeholders in activities and outreach actions: policymakers at International and national level; decision-makers in academia and industry; primary, secondary and high school students, teachers, and parents; university students at bachelor, master, and Ph.D. level; Academic stakeholders: young and senior researchers, Faculty at the assistant, associate and full professor level, Project managers, and HR experts; Industry stakeholders: IT industry directors, managers, engineers, and HR managers; the general public. It is important to involve more male colleagues on the activities and outputs, to build a more diverse and inclusive community.

3. Create a suitable framework to start a long-term collaboration of researchers. Plan activities to foster greater access and integration of female young academics with the main goals of working together to have more academic departments and companies implementing policies or adopting best practices to foster and support the careers of women in CS.

ACKNOWLEDGMENTS

We thank the whole Women in Informatics Research and Education (WIRE) group, especially the former leaders Erika Abraham, Lynda Hardman, and Jane Hillston. This work has been partially supported by the COST Action CA19122 - European Network for Gender Balance in Informatics and by NFR 295920 IDUN.

REFERENCES

[1] "She Figures 2018, European Commission." https://ec.europa.eu/info/publications/she-figures-2018_en (accessed Nov. 24, 2020).

[2] S. Zweben and B. Bizot, "Taulbee Survey Total Undergrad CS Enrollment Rises Again, but with Fewer New Majors; Doctoral Degree Production Recovers From Last Year's Dip," 2020.

[3] L. L. Wang, G. Stanovsky, L. Weihs, and O. Etzioni, "Gender trends in computer science authorship," Jun. 2019, doi: https://arxiv.org/abs/1906.07883.

[4] "Informatics Europe Higher Education Data Portal." https://www.informatics-europe.org/data/higher-education/ (accessed Dec. 09, 2020).

[5] L. J. Sax *et al.*, "Anatomy of an Enduring Gender Gap: The Evolution of Women's Participation in Computer Science," *J. Higher Educ.*, vol. 88, no. 2, pp. 258–293, Mar. 2017, doi: 10.1080/00221546.2016.1257306.

[6] "ACM Destinguished Speaker Professor Letizia Jaccheri." https://speakers.acm.org/speakers/jaccheri_10303 (accessed Nov. 24, 2020).

[7] "Informatics Europe Women in Informatics Research & Education (WIRE)." https://www.informatics-europe.org/activities/women-in-icst-research-and-education.html (accessed Nov. 24, 2020).

[8] "Gender equality in research and innovation | European Commission." https://ec.europa.eu/info/research-and-innovation/strategy/gender-equality-research-and-innovation_en (accessed Dec. 09, 2020).

[9] "Women in Digital. Digital Single Market Policy." https://ec.europa.eu/digital-single-market/en/women-ict (accessed Nov. 24, 2020).

[10] "Gendered innovations - Publications Office of the EU." https://op.europa.eu/en/publication-detail/-/publication/d15a85d6-cd2d-4fbc-b998-42e53a73a449 (accessed Dec. 06, 2020).

[11] "I'd blush if I could: closing gender divides in digital skills through education - UNESCO Digital Library." https://unesdoc.unesco.org/ark:/48223/pf0000367416 (accessed Nov. 24, 2020).

[12] E. Rubegni, M. Landoni, A. De Angeli, and L. Jaccheri, "Detecting Gender Stereotypes in Children Digital StoryTelling," in *Proceedings of the Interaction Design and Children - IDC '19*, 2019, pp. 386–393, doi: 10.1145/3311927.3323156.

[13] J. Gomez, L. Jaccheri, M. Maragoudakis, and K. Sharma, "Digital storytelling for good with Tappetina game," *Entertain. Comput.*, vol. 30, p. 100297, May 2019, doi: 10.1016/j.entcom.2019.100297.

[14] "The Girl Project Ada." https://www.ntnu.edu/girls (accessed Nov. 24, 2020).

[15] J. Brevik, J. C. T. Vidal, and L. Jaccheri, "Designing software to prevent child marriage globally," in *Proceedings of the Interaction Design and Children - IDC '19*, 2019, pp. 452–457, doi: 10.1145/3311927.3325322.

[16] I. O. Pappas, T. Aalberg, M. N. Giannakos, L. Jaccheri, P. Mikalef, and G. Sindre, "Gender Differences in Computer Science Education: Lessons Learnt from an Empirical Study at NTNU.," 2016.

[17] ICT specialists, "More than 8 million ICT specialists employed in the EU in 2016," Jul. 18, 2016. https://ec.europa.eu/eurostat/documents/2995521/8115840/9-18072017-AP-EN.pdf (accessed Nov. 24, 2020).

[18] I. Specialists, "Statistics on hard-to-fill vacancies in enterprises - Statistics Explained." https://ec.europa.eu/eurostat/statistics-explained/index.php/ICT_specialists_-_statistics_on_hard-to-fill_vacancies_in_enterprises (accessed Nov. 24, 2020).

[19] "Bamberg CS30 Strategy." https://nachwuchs.wiai.uni-bamberg.de/ (accessed Nov. 24, 2020).

[20] D. Lillis and S. McKeever, "Minerva Award Submission: CS4All Initiative," 2018. [Online]. Available: https://www.informatics-europe.org.

[21] "Irène Curie Fellowship at TU Eindhoven." https://www.tue.nl/en/working-at-tue/scientific-staff/irene-curie-fellowship (accessed Nov. 25, 2020).

[22] "Genie | Gender Initiative for Excellence | Chalmers." https://www.chalmers.se/en/about-chalmers/Chalmers-for-a-sustainable-future/initiatives-for-gender-equality/gender-initiative-for-excellence (accessed Nov. 25, 2020).

[23] "IDUN project at NTNU – from PhD to professor." https://www.ntnu.edu/idun (accessed Nov. 25, 2020).

[24] "Harvey Mudd Graduates Highest-ever Percentage of Women Physics and Computer Science Majors | College News | Harvey Mudd College." https://www.hmc.edu/about-hmc/2018/05/15/harvey-mudd-graduates-highest-ever-percentage-of-women-physics-and-computer-science-majors/ (accessed Nov. 24, 2020).

[25] "CMU's Proportion of Undergraduate Women in Computer Science and Engineering Soars Above National Averages - News - Carnegie Mellon University." https://www.cmu.edu/news/stories/archives/2016/september/undergrad-women-engineering-computer-science.html (accessed Nov. 24, 2020).

[26] "Grace Hopper Celebration," 2021. https://ghc.anitab.org/ (accessed Nov. 24, 2020).

[27] "womENcourage 2020 | ACM," 2020. https://womencourage.acm.org/2020/ (accessed Nov. 24, 2020).

[28] "IEEE WIE International Leadership Conference," 2020. http://ieee-wie-ilc.org/ (accessed Nov. 24, 2020).

[29] "Association for Computing Machinery | Supporting, celebrating and advocating for Women in Computing." https://acmweurope.acm.org/ (accessed Nov. 24, 2020).

[30] "European Centre for Women and Technology (ECWT)." http://www.ecwt.eu/en/home (accessed Nov. 24, 2020).

[31] "Cyprus Computer Society » Project » CEPIS Women In ICT Expert Group." https://ccs.org.cy/en/projects/cepis-women-in-ict-expert-group-13 (accessed Nov. 24, 2020).

[32] "Global Partnership for Gender Equality in the Digital Age." https://www.itu.int/en/equals/Pages/default.aspx (accessed Nov. 24, 2020).

[33] "Women4IT is a project aimed at developing innovative solutions to increase the numbers of vulnerable girls and young women in the European digital economy." https://women4it.eu/ (accessed Nov. 24, 2020).

[34] "COST Action CA19122 EUGAIN European Network For Gender." https://www.cost.eu/actions/CA19122 (accessed Nov. 24, 2020).

[35] "LIBRA | Unifying innovative efforts of European research centres to achieve gender equality in academia." https://www.eu-libra.eu/ (accessed Dec. 09, 2020).

[36] "Geecco : Home." http://www.geecco-project.eu/home/ (accessed Dec. 09, 2020).

[37] "Plotina - Promoting gender balance and inclusion in research, innovation and training." https://www.plotina.eu/ (accessed Dec. 09, 2020).

[38] "Gender Action." https://genderaction.eu/ (accessed Dec. 09, 2020).

[39] "Startseite - FG Frauen und Informatik - Gesellschaft für Informatik e.V." https://fg-frauen-informatik.gi.de/ (accessed Dec. 09, 2020).

[40] "EQUAL-IST | Gender Equality." https://equal-ist.eu/ (accessed Nov. 24, 2020).

[41] Elsevier, "The researcher journey through a gender lens." https://www.elsevier.com/research-intelligence/resource-library/gender-report-2020 (accessed Nov. 24, 2020).

[42] "Gender Equality in Academia and Research - GEAR tool | EIGE." https://eige.europa.eu/gender-mainstreaming/toolkits/gear (accessed Nov. 24, 2020).